\documentstyle[eqsecnum,preprint,aps,prb,amssymb,psfig]{revtex}
\begin{document}
\preprint{}

\title{Transverse ordering of an antiferromagnet in a field with
  oblique angle to the easy axis}

\author{M.~Acharyya \cite{muktishEmail}, U.~Nowak \cite{uliEmail} and
  K.~D.~Usadel \cite{usadelEmail}}

\address{Theoretische Tieftemperaturphysik,
  Gerhard-Mercator-Universit\"{a}t-Duisburg, 47048 Duisburg/
  Germany\\}

\date{\today}

\maketitle

\begin{abstract}
  Motivated by the recent experimental observations [Phys. Rev. B {\bf
    57} R11051 (1998)] of transverse spin ordering in ${\rm FeBr_2}$
  induced by a magnetic field with oblique angle to the easy axis of
  the system, we performed extensive Monte Carlo simulations of a
  classical anisotropic Heisenberg model. We have calculated the
  specific heat and the parallel and perpendicular components of the
  magnetisation as well as the antiferromagnetic order parameter
  and studied these quantities
as a function of temperature.  A tilted spin-flop phase
  is obtained for certain parameter values.  Many of the effects
  occurring in connection with this phase agree qualitatively well
  with the experimental facts.
\end{abstract}

\section{Introduction}
Among many other magnetic materials, so called {\it metamagnets} show
interesting phase transitions induced by an external magnetic field
\cite{review}. Especially, the multicritical behavior in the
field-temperature-plane of the phase diagram is subject of immense
interest. In recent experiments \cite{azev,katori,oleg} the metamagnet
${\rm FeBr_2}$ was studied.  Cooling the sample in zero field the
well-known \cite{review} transition from the paramagnetic to the
antiferromagnetic state leading to a divergence of the specific heat
at the respective N\'eel temperature was observed.  For finite applied
magnetic fields (along the crystallographic c-axis) the magnetic part
of the specific heat of this system showed a peculiar shape.  As the
field increases, the specific heat develops an anomalous peak 
(the structure containing a broad noncritical anomaly at $H_{-}(T)$,
and a sharp peak at $H_{1}(T)$, where $H_{-} < H_1 < H_c$ \cite{oleg})
at a
temperature lower than the corresponding critical temperature. This
anomalous peak may indicate an additional second phase transition
besides the usual transition from the paramagnetic (saturated) to the
antiferromagnetic phase. To identify the nature of a possible third
phase is the main goal of both, theoretical
\cite{selke1,selke2,pleim} and experimental work \cite{katori,oleg}.

For a simple and qualitative understanding, Monte Carlo simulations
have been performed for an Ising model \cite{selke1}.  Considering the
hexagonal lattice structure of ${\rm FeBr_2}$, ferromagnetic
intra-planar interaction and antiferromagnetic inter-planar
interactions, it has been shown \cite{selke1} that the anomalous peak
of the specific heat (at $H_{-}(T)$) can be reproduced with interaction parameters
obtained from spin wave analysis and neutron scattering experiments.
The "phase boundary" obtained from Monte Carlo simulations agreed
qualitatively well with the experimental one \cite{selke1}. It has
been conjectured \cite{selke1} that the anomaly line is the "border"
between antiferromagnetic at low temperature and a "mixed phase", where it
was speculated that due to the positive axial field small clusters of
positive spins in the negative sea may form a stable phase.  The
detailed characterization of this "intermediate phase" is missing in the
literature.

However, the recent experimental observations \cite{oleg} of a
transverse spin ordering associated with a weak first order transition
(at $H_1(T)$)
and a sharp peak of the specific heat cannot be explained by a simple
Ising model.  A model with transverse spin components is necessary.
A disorder-order transition of the $m_s=0$ spin components probably due
to off-diagonal exchange \cite{mukamel} was conjectured \cite{oleg}.
Motivated by this conjecture, the so-called
semi-classical Heisenberg model including off-diagonal exchange
interactions has been studied recently \cite{selke2} by Monte Carlo
simulation.  In this model, the axial component of the spin vector is
quantized (it can take values -1, 0 and +1) while the planar component
is a classical vector which can rotate continuously in the transverse
plane. One can consider this model to be a (de)coupled combination of
a $S=1$ Ising model with a kind of classical XY- model and
consequently, with Ising-like anisotropy,
one observes always two sharp peaks in the specific heat
even at zero axial field (surprisingly also with ferromagnetic interaction 
and no off-diagonal exchange interaction) \cite{pleim}. 
The appearance of these two peaks at zero field is in contradiction to
 the experimental evidence of a critical end point
on the anomaly line at nonzero axial field (see phase diagram of ref \cite{oleg}).
Also, the sequence of
the different orderings (planar and axial) with temperature seems to be
reversed \cite{selke2,pleim} as compared with the experimental facts
\cite{oleg,neutron}. The microscopic description of the spin
configuration in different phases has not been worked out so far.

These shortcomings of the semi-classical model led us to search for
a different approach. We found that a much simpler model, namely a classical
Heisenberg model can explain some of the recent experimental facts.
In our paper, we report on our results
from Monte Carlo simulations of an anisotropic classical Heisenberg
model in the presence of a magnetic field where the field may have an
oblique angle to the easy axis of the system. We study the temperature
variations of the specific heat, the transverse and axial
magnetisations and antiferromagnetic order parameters and compare
directly with experimental observations \cite{oleg,neutron}. 
We are especially interested in the nature
(microscopic configuration) of the phase in between the critical line
and the so-called anomaly line of the phase diagram of ${\rm FeBr_2}$
\cite{azev}. 
Our
results show quite close resemblance to the recent experimental facts
\cite{oleg,neutron}. 
 The paper is organized as follows: in the next section
we present the model, in section (III) the Monte Carlo simulation
scheme is discussed, section (IV) contains the simulational results
and the comparison with experimental facts. At the end of this section,
the microscopic spin configuration in different phases is shown.
The paper ends with a summary and concluding remarks in section V.

\section{The classical anisotropic Heisenberg Model}
     
The classical, anisotropic Heisenberg model with competing interactions
in presence of a magnetic field can be represented by the 
Hamiltonian
\begin{eqnarray}
  \cal H = & - & J \sum_{\langle ij \rangle} \vec S_i \cdot \vec S_j -
  J' \sum_{\langle ij \rangle'} \vec S_i \cdot \vec S_j \nonumber \\ &- &
  D\sum_i (S_i^z)^2 -  \vec H \cdot \sum_i \vec S_i,
\end{eqnarray}
\noindent where $\vec S_i$ represents a classical spin vector of magnitude unity
at site $i$ of the lattice. This spin vector may point into any
direction in spin space continuously.  For simplicity, we have chosen
a tetragonal lattice of linear size $L$. The ranges of interactions are
limited to the nearest neighbors only where the first sum is over the
intra-planar exchange interactions which are ferromagnetic ($J > 0$)
and the second sum is over the inter-planar exchange interactions
which are antiferromagnetic ($J' < 0$).  $D$ is the uniaxial
anisotropy constant favoring the spin to be aligned either parallel or
antiparallel to the $z$-axis and $\vec H$ is the external, uniform
magnetic field.  We use periodic boundary conditions in all
directions.

We have performed Monte Carlo simulations for the system described
above where we used a system size of $L = 20$. Measuring all
energetical quantities in units of the ferromagnetical intra-planar
interaction $J$ we set the antiferromagnetic inter-planar interaction
to $J' = -0.5J$) and the anisotropy to $D=0.3J$. It should be noted
here that very large values of $D$ will yield Ising-like behavior. In
order to be able to observe transverse ordering one has to choose
lower values for the anisotropy. We have suitably chosen the parameter
values in such a way that the anisotropy is high enough to yield a
longitudinal antiferromagnetic phase at low magnetic field and low
enough to allow for reasonably large transverse spin components so
that the qualitative behavior of the transverse components of
magnetisation and order parameter can be observed within the Monte
Carlo method. The specific choice of the parameter values was
optimized by trial-and-error. We are aware of the fact that our choice
of parameters is not realistic compared to ${\rm FeBr_2}$. Especially,
the value of $D$ is much too low in our simulations. 
On the other hand, it is known that the exchange interaction takes
place between a large number of spins, it is not restricted to the
nearest neighbours only.
The
transverse ordering in experimental systems is much smaller compared
with the longitudinal one (less than 1\% \cite{neutron}). These effects
are too small to be observabed in a realistic, quantitative
simulation. Hence, we restrict ourselves to a pure qualitative
description of certain effects that might be comparable to those found
experimentally.

 \bigskip

\section{Monte Carlo Simulation scheme}

We performed extensive Monte Carlo simulations of the system above
using the following algorithm. At fixed temperature $T$ and field
$\vec H$, we choose a lattice site $i$ randomly and updated the spin
value $\vec S_i$ to $\vec S'_i$ (randomly chosen on an unit sphere) by
using the Metropolis rate \cite{binder}
\[
W(\vec S_i \to \vec S'_i) = {\rm Min}[1,\exp(-\Delta {\cal H} / k_B T)]
\]
\noindent where $\Delta \cal H$ is the change of energy due to the
change of the direction of the spin vector from $\vec S_i$ to $\vec
S'_i$.  We set the Boltzmann constant to $k_B = 1$. $L^3$ such random
updates of spins is defined as one Monte Carlo Step per site (MCSS).

Starting from an initially random configuration (corresponding to a
high temperature phase) we equilibriate the system up to $4\times10^4$
MCSS and calculated thermal averages and fluctuations from further
$4\times10^4$ MCSS. Hence, the total length of the simulation for one
fixed temperature $T$ is $8\times10^4$ MCSS. Then we decrease the
temperature and use the last spin configuration of the simulation
at this temperature
 as the initial configuration for the calculation for the next
temperature. In this way we simulate a cooling procedure which is
closer to equilibrium compared to starting at each temperature with a
random spin configuration.  The CPU time needed for $8\times10^4$ MCSS
is approximately 1 hour on an IBM RS/6000-590 workstation.

We have calculated the following quantities:
\begin{enumerate}
\item Sublattice magnetisation components for odd and even labelled
  planes:
  \[
  m^q_{o,e} = \frac{2}{L^3} \sum_{i \in \{e\},\{o\}} \left\langle
  S^q_i \right\rangle 
  \]
  where $q \in \{x,y,z\}$ and the sum is over all sites in either even
  or odd labeled planes. $<\ldots>$ denotes an average over time
  (MCSS) (assuming ergodicity and, hence, that an ensemble average and
  the time average yields the same results).
\item Longitudinal antiferromagnetic order parameter:
  \[
  O_{AF}^z = \frac{1}{2} |(m^z_o-m^z_e)|
  \]
\item Longitudinal ferromagnetic order parameter:
  \[
  M_{F}^z = \frac{1}{2} (m^z_o+m^z_e)
  \]
\item Transverse antiferromagnetic order parameter:
  \[
  O_{AF}^{xy} = \frac{1}{2} {\sqrt {(m^x_o-m^x_e)^2 + (m^y_o-m^y_e)^2}}
  \]
\item Transverse ferromagnetic order parameter:
  \[
  M_{F}^{xy} = \frac{1}{2} {\sqrt{(m^x_o+m^x_e)^2 + (m^y_o+m^y_e)^2}}
  \]
\item Total energy per lattice site
  \[
  E = \frac{1}{L^3} \langle \cal H \rangle
  \]
\item Specific heat per site:
  \[
  C = L^3 \delta E^2/(k_B T^2)
  \]
\noindent where, $\delta E^2 = \langle \frac{1}{L^6} {\cal H}^2 \rangle -
\langle\frac{1}{L^3} {\cal H} \rangle^2$ are the fluctuations of the
energy. 

\end{enumerate}

Note that the specific heat $C$ can also be obtained from the
temperature derivative of the energy, $dE/dT$. Interestingly, it turns
out to be a criterion for equilibrium that the two definitions of $C$
are identical during our simulations.

\bigskip

\section{Numerical Results}

As a primary check, we show in Fig. \ref{f:h=0} the temperature
variation of the longitudinal antiferromagnetic order parameter,
$O^z_{AF}$, the transverse antiferromagnetic order parameter,
$O^{xy}_{AF}$, and the magnetic specific heat, $C$, at zero field,
$\vec H = 0$.

Our results indicate that at zero field only one transition is observed from a
paramagnetic to an antiferromagnetic state where the spins of odd and
even planes are aligned alternate parallel and antiparallel to the
z-axis (Fig.\ref{f:h=0}a). The transverse antiferromagnetic and
ferromagnetic order parameters remain zero for all temperatures.
Consequently the temperature variation of the specific heat (Fig.
\ref{f:h=0}b) shows one single peak at the N\'eel temperature $T_N
\cong 1.28$. This is also observed in experiments as a well-known fact
\cite{katori}. It should be emphasized here that the semi-classical
Heisenberg model with Ising-type anisotropy shows two peaks following
from two transitions for zero field (see Fig. 1 of Ref. \cite{pleim}),
which is not consistent with the experimental facts.

In an applied field parallel to the easy axis the (longitudinal)
antiferromagnetic ordering is stable for fields up to $H_z \leq 0.64$.
The peak position of the specific heat shifts towards lower
temperature as one increases the axial field $H_z$.  This result is
also consistent with the experimental observations \cite{katori,oleg}.

To compare with recent experimental observations \cite{oleg}, we now
apply a small transverse field ($H_x=0.1$) in addition to an axial
field of $H_z=0.7$. It should be noted that in real experiments
\cite{oleg} the effect of a transverse field has been incorporated
just by tilting the sample by a certain angle $\theta$ with respect to
the direction of the field.  Fig. \ref{f:c-and-e} a shows the
temperature variation of the magnetic specific heat measured from
both, the fluctuations of the energy and the temperature derivative of
the energy.  Both results agree reasonably well and show two peaks in
agreement with experimental facts \cite{oleg}. The high temperature
peak is usually called \cite{oleg} the critical one while the low
temperature sharp peak close to the broad anomalous maximum of the
specific heat (not reproduced in our simulations)
is not yet explained. For a direct comparison we refer to
see Fig. 2 of Ref.  \cite{oleg}, keeping in mind that in our simulation
we fixed the
field and varied the temperature whereas the reverse was done in
experiments \cite{oleg}.

The low temperature sharp peak can be identified as
signature of a first order phase transition while the high temperature
peak seems to be associated with a second order phase transition.
This follows immediately from the temperature variation of the total
energy $E$ which is presented in Fig. \ref{f:c-and-e}b. At low
temperature there is a jump of the energy - a latent heat - which
appears as a sharp peak in the specific heat. To characterize the
nature of this phase in the intermediate temperature range (in between
the two peaks of the specific heat) we have studied also the
temperature variation of the longitudinal and transverse order
parameters, respectively.

Fig. \ref{f:o-and-m}a shows the temperature variation of the
longitudinal antiferromagnetic order parameter $O^z_{AF}$. The
behavior of $O^z_{AF}$ clearly indicates two phase transitions, one at
higher temperature ($T \sim 1.0$) which is continuous and a second one
which is of first order (or discontinuous) at lower temperature ($T
\sim 0.78$).  The temperature variations of the longitudinal
ferromagnetic order parameter ($M^z$) and the transverse
antiferromagnetic order parameter ($O^{xy}_{AF}$) are shown in Fig.
\ref{f:o-and-m}b.  The transverse antiferromagnetic spin ordering is
evident in the intermediate range of temperature. This result is very
similar to recent experimental \cite{neutron} observations made by
neutron diffraction.

We conclude that during cooling from high temperatures, the system
first orders continuously to a transverse antiferromagnetic phase.
The corresponding ordering temperature is marked as $T_c$. This
transverse antiferromagnetic order increases as the temperature
decreases and at lower temperature a second transition occurs where
the transverse antiferromagnetic order jumps to a lower value leading
to a mainly longitudinal antiferromagnetic order. In other words, this
second transition corresponds to a discontinuous rotation of the
staggered magnetisation vector from a mainly transverse direction to a
mainly longitudinal one. It should be mentioned here that the opposite
scenario was observed in the semi-classical model with off-diagonal
interaction studied recently (see Fig. 9 of Ref.  \cite{pleim}).

For a direct comparison with the earlier experiments \cite{oleg}, we
have calculated the magnetisation components parallel
($M_{\parallel}$) and perpendicular ($M_{\perp}$) to the total applied
field $\vec H = H_x \hat x + H_z \hat z$ from the longitudinal and
transverse magnetisation components.  In experiments, the latter are
termed as $M_{ax}$ and $M_{pl}$, respectively. We have, $\theta =
\tan^{-1}(H_x/H_Z) \approx 8.2^\circ$.  In the experiment \cite{oleg}
this tilting angle was even larger (approximately $30^\circ$) but our
choice for this angle $\theta$ is restricted by the parameter values
used in the simulation.  $M_{\parallel}$ and $M_{\perp}$ can be
readily calculated just by applying a rotation of angle $\theta$ which
yields $M_{\parallel}=M^z_F\cos \theta + M^{xy}_F\sin \theta$ and
$M_{\perp}=-M^z_F\sin \theta + M^{xy}_F\cos \theta$. The temperature
variations of $M_{\parallel}$ and $M_{\perp}$ obtained in this way are
shown in Fig.\ref{f:o-and-m}c. The weak first order jump is evident
and the data agree qualitatively with the experimental diagram (see
Fig. 3 of Ref. \cite{oleg}). The transition at higher temperature is
indicated by a marker $T_c$, where the slope of $M_{\parallel}$ (i.e.,
$dM_{\parallel}/dT$) becomes maximal.

What will be the microscopic spin structure in all different phases?
The high temperature phase is disordered, of course with a
paramagnetic response to the external field. Hence, as the temperature
decreases the longitudinal component of total magnetisation increases.
At $T_c$, the transverse antiferromagnetic order starts to develop and
consequently, the longitudinal component of the total magnetisation
decreases. The spin structure of this phase is sketched in Fig. \ref{f:spins}
(marked as TSF). It is a spin-flop (SF) phase, 
slightly tilted along the positive $x$-direction due to
presence of the transverse field.  We call it a tilted spin-flop phase
(TSF).

To understand this phase let us first recall the structure of a spin-flop
phase. In a pure spin-flop phase (drawn and marked as SF in Fig. \ref{f:spins}), one finds
longitudinal ferromagnetic order
and transverse antiferromagnetic order as follows from the $x$- and
$z$-components of the spin vector which are also shown. 
Lowering the temperature from a paramagnetic phase, first the
longitudinal magnetisation will increase and then will remain constant
if the angle between two spins remains constant or increases if the
angle between two spins decreases.  At the transition temperature
$T_c$ the slope $dM^z_{AF}/dT$ will change rapidly.  The longitudinal
antiferromagnetic order parameter remains zero since one has equal
values of $m^z_o$ and $m^z_e$. One can characterize the spin-flop (SF)
phase by, $M^z_F \neq 0$, $O^{xy}_{AF} \neq 0$, $O^z_{AF} = 0$ and
$M^{xy}_F = 0$.  It is mainly a coexistence of axial ferromagnetic
order and transverse antiferromagnetic order.

However, in the tilted spin-flop phase, i. e. in presence of a
transverse field, the spins in one layer will be 
more aligned along the positive $x$-direction compared to the 
spins in the neighbouring layer (see TSF in Fig. \ref{f:spins}). 
This will increase the angle between the two spins and as a
result, the longitudinal magnetisation will start to decrease as one
decreases the temperature.  Almost the same effect can be observed in
the temperature variation of $M_{\parallel}$ (see our Fig.
\ref{f:o-and-m}c and for comparison also the experimental situation,
Fig. 3 of Ref.\cite{oleg}).  Due to unequal values of $m^z_o$ and
$m^z_e$ one obviously will find nonzero values of the longitudinal
antiferromagnetic order parameter in the TSF phase (see Fig.
\ref{f:o-and-m}a).  But nevertheless the system is effectively
ferromagnetically ordered since the signs of the values of $m^z_e$ and
$m^z_o$ are the same even when the absolute values are different so
that the longitudinal antiferromagnetic order parameter is nonzero in
this phase. Since the absolute values of the transverse
magnetisations of the two different sublattices are different
(although they are oppositely directed), the transverse magnetisation
is nonzero.  This observation has also been made in experiments
\cite{neutron}.  Hence in the TSF phase it is $M^z_F \neq 0$,
$O^{xy}_{AF} \neq 0$, $O^z_{AF} \neq 0$ and $M^{xy}_F \neq 0$.

After a further decrease of temperature one will encounter a phase
with longitudinal antiferromagnetic (AF) order.  The transition from
TSF- to AF-phase is of first order. This is consistent with the
experimental observations \cite{oleg,neutron}.  The weak jumps of
$M_{\parallel}$ and $M_{\perp}$ (see our Fig.  \ref{f:o-and-m}c and
for comparison with experiments Fig. 3 of Ref.\cite{oleg}) is a
signature of a discontinuous transition from a tilted spin-flop (TSF)
phase to a longitudinal antiferromagnetic (AF) phase.  In a pure
longitudinal antiferromagnetic (AF) phase, $M^z_F = 0$, $O^{xy}_{AF} =
0$, $O^z_{AF} \neq 0$ and $M^{xy}_F = 0$. Strictly speaking, due to the
application of a small $H_x$ one will have very small but nonzero value of
$M^{xy}_F$.


In addition, we have also studied the temperature variation of the
transverse antiferromagnetic susceptibility ($\chi^{xy}_{AF} = {{L^3
    (\delta O^{xy}_{AF})^2} \over {k_B T}}$) shown in Fig.
\ref{f:chi}. The two transitions, i.e., at high temperature from saturated
paramagnetic to tilted spin-flop and at low temperature from a tilted spin-flop to
longitudinal antiferromagnetic phase, are evident from the figure.

\bigskip

\section{Summary}

Motivated by recent experimental observations \cite{oleg} in the
metamagnet ${\rm FeBr_2}$, we have studied a classical anisotropic
Heisenberg model with ferromagnetic intra-planar interaction and
antiferromagnetic inter-planar interaction by Monte Carlo simulations.
We focused on the temperature variations of the magnetic specific
heat, longitudinal and transverse order parameters (both ferromagnetic
and antiferromagnetic) and $M_{\parallel}$ and $M_{\perp}$, where the
system is in a magnetic field tilted to the easy axis of the system. 

Transverse spin ordering and a weak first order transition (additional
to the well known antiferromagnetic transition) associated with a very
sharp peak of the magnetic specific heat at low temperature are
observed in agreement with experiments \cite{oleg,neutron}. The high
temperature phase transition is identified as a continuous transition
from a paramagnetic phase to a tilted spin-flop phase while the low
temperature transition is discontinuous and from tilted spin-flop
phase to a longitudinal antiferromagnetic phase.

None of the models studied so far theoretically can provide a
reasonably well explanation for all experimental facts observed in the
${\rm FeBr_2}$ metamagnet at the same time.  Monte Carlo calculations in
an Ising model \cite{selke1} on a hexagonal lattice with realistic
interaction parameters can reproduce the broad anomalous maximum of
the specific heat at $H_{-}(T)$. This anomaly is not reproduced within our
simulations. It was shown \cite{selke1,selke2,pleim} that this anomaly 
is due to a strong Ising character
of ${\rm FeBr_2}$ and it is due to the fact that one needs a large
number of interlayer interaction neighbors.
 
On the other hand, recent experimental observations of transverse ordering
\cite{oleg} cannot be explained by an Ising model \cite{selke1}. The
semi-classical Heisenberg model with off-diagonal interaction
\cite{selke2,pleim} contains the anomaly of the specific heat as well
as the sharp peak additional to the usual transition. But the sequence
of different ordering seems to be in contradiction with recent neutron
diffraction results \cite{neutron}.  Most importantly, it gives two
transitions (associated with two peaks of the specific heat) even in
zero field. 
However, very probably, \cite{katori,oleg} the phase line $H_1(T)$ ends
up at a critical end point at nonzero field.
 On the other hand, our much simpler approach, with a classical
anisotropic Heisenberg model, can explain some of the recent
experimental facts \cite{oleg,neutron} and it can also provide a
microscopic description of the different ordering.  To find a model which can
explain the entire phase diagram of the ${\rm FeBr_2}$, in our opinion
is still an open problem.

\section*{Acknowledgments}
We would like to thank W. \ Kleemann, O.
\ Petracic and Ch. \ Binek for important discussions at various stages
during the development of this work and for making us their
Neutron diffraction data available prior to publication. We thank
W. \ Kleemann for critical reading
of our manuscript and for various comments on it.
Finally, we thank W.\ Selke and M.\ Pleimling for discussions and comments.
The financial support from
Graduiertenkolleg "Struktur und Dynamik heterogener Systeme" at the
University of Duisburg is gratefully acknowledged.

\begin{figure}

\caption{Temperature variations of (a) longitudinal ($O^z_{AF}$)
  and transverse ($O^{xy}_{AF}$) antiferromagnetic order parameter
  (solid line is just connecting the data points) and (b) specific
  heat $C$ (solid line represents ${dE \over dT}$). $\vec H = 0$.}
\label{f:h=0} \end{figure}

\begin{figure}
\caption
{Temperature variations of (a) specific heat $C$ (the continuous line
  represents $dE/dT$) and (b) total energy $E$, for $H_z=0.7$ and
  $H_x=0.1$.}
\label{f:c-and-e}
\end{figure}

\begin{figure}
\caption
{Temperature variations of (a) longitudinal antiferromagnetic order
  parameter ($O^z_{AF}$), (b) longitudinal ferromagnetic ($M_z$) and
  transverse antiferromagnetic ($O^{xy}_{AF}$) order parameter, and
  (c) $M_{\parallel}$ and $M_{\perp}$ as explained in the text.  Solid
  lines in (a) and (b) are just connecting the data points. $H_z=0.7$
  and $H_x=0.1$.}

\label{f:o-and-m}
\end{figure}

\begin{figure}
\caption
{Schematic representation of an antiferromagnetic
  (AF) phase, a spin-flop (SF) phase, and a tilted spin-flop (TSF)
  phase. Each vector may represents the magnetisation of one plane of the
  system.}
\label{f:spins}
\end{figure}

\begin{figure}
\caption
{Temperature variation of the transverse antiferromagnetic
  susceptibility ($\chi^{xy}_{AF}$) for $H_x=0.1$ and $H_z=0.7$.
  The solid line is just connecting the data points.}
\label{f:chi}
\end{figure}

\end{document}